\documentclass[doublespacing]{elsart}
\usepackage{amssymb}
\usepackage{graphics}
\usepackage{epsfig}

\begin{document}

\begin{frontmatter}
\title{Amplitude scaling behavior of band center
states of Frenkel exciton chains with correlated off-diagonal disorder}
\author{Ibrahim Avgin}
\address{Department of Electrical and Electronics
Engineering, Ege University,
Bornova 35100, Izmir, Turkey}

\author{David L Huber}\footnote{Corresponding author: e-mail address
dhuber@mail.src.wisc.edu, Tel. No. (608) 238-8880, Fax. No. (608)
265-2334}  \\
\address{Department of Physics, University of Wisconsin-Madison,
Madison, Wisconsin 53706 USA}


\begin{abstract}
We report the amplitude scaling behavior of Frenkel exciton chains with 
nearest-neighbor correlated off-diagonal random interactions.  The band 
center spectrum and its localization properties are investigated through 
the integrated density of states and the inverse localization length.  The 
correlated random interactions are produced through a binary sequence 
similar to the interactions in spin glass chains.  We produced sets of data 
with different interaction strength and "wrong" sign concentrations that 
collapsed after scaling to the predictions of a theory developed earlier 
for Dirac fermions with random-varying mass.  We found good agreement as 
the energy approaches the band center for a wide range of concentrations. 
We have also established the concentration dependence of the lowest order 
expansion coefficient of the scaling amplitudes for the correlated 
case.  The correlation causes unusual behavior of the spectra, 
i.e., deviations from the Dyson-type singularity.
\end{abstract}

\end{frontmatter}


\section{Introduction}
Low-dimensional disordered excitonic systems have generated renewed interest
recently. For example, the possibility of using optically active polymers in
various photonic applications such as flat panel displays and light emitting
diodes has led to increased interest in their electronic and optical
properties \cite{orel,jpc,cpa}. 
We study here numerically the dynamics of Frenkel excitons near the band
center in a system with nearest-neighbor off-diagonal random interactions.
This study is, we believe, the first to treat correlated disorder in the
off-diagonal dipolar interactions in an exciton system which focuses on the
behavior near band center $E=0$. However, an analysis of a similar problem
in the correlated electronic system has appeared \cite{irn} but the authors
considered correlations those were different from ours, and the interactions
were not of the dipolar type \cite{av99}. Many efforts have also been spent
on the same off-diagonal tight-binding system using supersymmetric methods
(SUSY) \cite{fisher} where the interactions are formally similar to our
dipolar interactions but the authors eventually worked with continuous
variables whereas ours are discrete. Later, correlated disorder, in
particular the exponential type, was included for the same tight binding
model using the SUSY \cite{jp1,jp2}. Our results can thus be compared with what
has been developed for the SUSY methods.  
This study is 
important because most realistic random disordered systems 
have nonlocal correlations.  

The nature of the excitations at the band center of random off-diagonal
exciton models has been discussed for a long time, and it has been found
that the spectrum has a singularity of the Dyson type \cite{dyson53} in
which spectral properties in the vicinity of the band center are functions
of $\ln (E)$ where $E$ is the energy relative to the center of the band.
The question of the band center localization has also been argued at
length, and it was claimed that the band center mode may be weakly localized
because of the strong fluctuations \cite{lic7, delyon83,av98}.
Ziman \cite{ziman7,eggarter} analyzing various uncorrelated distributions of
the random coupling found the low-energy behavior of the integrated density
of states ($IDOS$)
 
\begin{equation}  \label{idos}
IDOS-{\frac{1 }{2}}=\frac {V^U} {{\vert ln(E^{2}) \vert}^{2}},
\end{equation}

\noindent and the inverse localization length ($ILL$)

\begin{equation}  \label{ill}
ILL=\frac {V^U} {\vert ln(E^{2}) \vert},
\end{equation}

\noindent in which $E$ is\ the energy and the logarithmic variance 
\cite {ziman7,eggarter} $V^{U}={\frac{1}{2}\left[ {\langle (\ln
(J^{2}))^{2}\rangle }-{\langle \ln (J^{2})\rangle }^{2}\right] }$ is the
amplitude factor, $J$ being the off-diagonal coupling specified below. For
definitions of the $ILL$ and the $IDOS$ see Ziman's work \cite{ziman7}
particularly his discussion after the equation of the motion. Notice that
the averages above are independent of the lattice site $n$ since
uncorrelated random variables are involved; otherwise they are site
dependent. Concerning the shape of the $IDOS$ and the $ILL$, according to
Ziman's analysis \cite{ziman7}, they should depend on disorder only through
a multiplicative constant, the variance in the distribution of $ln(J^{2})$.
As a result, the spectra near $E = 0$ have the same form
for all sets of independent, identically distributed random variables $J$.
In this context, such a property is referred to as scaling.
When there is scaling, the data can be collapsed onto a
single curve by dividing through by a factor that depends on the properties
of the disorder but is independent of the energy. Although the above
mentioned results were developed for uncorrelated disorder, our analysis
shows that qualitatively similar scaling
behavior for limited range of concentrations
and energy can be obtained for a system with
exponential correlations but with a modified amplitude factor $V^{C}$.
Moreover, for a certain range of correlation lengths a higher order
expansion of the Dyson singularity can reproduce the numerical data better
than just a single term.

A disordered 1D exciton model with nearest-neighbor interactions has the
following equation of motion

\begin{equation}  \label{tb}
J_{n+1}U_{n+1} + J_{n-1}U_{n-1} = EU_{n},
\end{equation}

\noindent where $U_{n}$ denotes the usual exciton creation operator and $
J_{n}$ are the couplings. This is equivalent to the XY model in a strong
applied field in the $z$-direction ($<Sz>=S$). 
In our numerics we calculated
iteratively the Lyapunov exponent 
\cite{ziman7,eggarter} 
$\gamma (E)=\frac{1}{N}\ln {\frac{U_{N}}{ U_{1}}}$ (see appendix), 
where $N$ is the number of
sites \cite{ziman7,eggarter}, 
whose real part is related to the $ILL$ while
the imaginary is related to the $IDOS$. We consider couplings that have
correlated or uncorrelated (for comparison) disorder and are of a dipolar
type. Since the displacements of 
the atoms in quasi-one dimensional systems
are likely to be small compared to the lattice constants, it is usually
enough to consider the first order term in the expansion of $J$ in terms of
the displacements \cite{dip}. 
For the first order expanded coupling we have $
J_{n}=1+A\xi _{n}$ with unit 
lattice constant where $A$ is the strength of
the disorder and ${\xi _{n}}$ is a random sign variable that is
exponentially correlated $<\xi _{n}\xi _{m}>=e^{-\frac{|m-n|}{l(c)}}$ 
where $l(c)$ is the correlation length as a function of 
"wrong sign" concentration $c$ (see below). 
Thus the dipolar coupling $J_n$ has a non
fluctuating part and a fluctuating part as in Dirac fermions with
random-varying mass \cite{jp1,jp2}.

The generation of random binary sequences with correlation was deviced 
by implementing the convolution method of constructing a random squence. 
Such squence can be used in various applications as in designing 
low-dimensional devices with desired properties 
\cite{jp1,izr99}.  A detailed discussion on this topic recently
appeared \cite{izr07}.  However, a more
efficient and easily handled correlated sequence, related to the spin glass
chain problem \cite{avgin93, boukahil89}, can be obtained through
uncorrelated random numbers as outlined below. The correlated random number
at site $n$ follows the relation $\xi _{n}=\xi
_{n-1}x_{n}=x_{1}x_{2}x_{3}\ldots x_{n-1}x_{n}$ 
with the distribution (of $x_{i}$ $i$ within $[1,n]$)

\begin{equation}
P(x_{i})=(1-c)\delta (x_{i}-1)+c\delta (x_{i}+1) ,  \label{dist}
\end{equation}

\noindent The $x_{i}$ are uncorrelated between different sites and $c$ is
the concentration of "wrong signs" such that $c=0$ and $c=1$ mean no
disorder. Clearly $\xi _{n}$ is exponentially correlated $\langle \xi
_{n}\xi _{m}\rangle =(1-2c)^{|m-n|}$ where one can define a correlation
length \cite{vulp89} $l(c)=-1/ln|1-2c|$. For the numerical computation
uncorrelated $x_{i}$ are obtained from the uniformly distributed random
number generator sequence. For the uncorrelated results we set $\xi
_{n}=x_{n}$ at the particular concentration $c=0.5$ where both correlated
and uncorrelated cases produce identical results since $l(0.5)=0$.

It should be emphasized that while the model for uncorrelated disorder
and the model for correlated disorder involve the parameter $c$,
in the latter model $c$ determines both the fraction of ‘
wrong sign’ bonds and the correlation length. The two models ‘
intersect’ at $c = 0.5$ where there are equal numbers of
$\pm$ interactions and the correlation length is $0$.
We have carried out extensive numerical studies to verify the
equation for the correlation length given above.

We note here that unlike the off-diagonal case above, 
many works have recently appeared on the
diagonal tight binding model with correlated disorder dealing with various
aspects of the problem \cite{yan05,tit03,deyc03}. It was revealed that even
short range correlated disorder at the band edge caused the anomalies in the
spectrum \cite{scho05}. Frenkel excitons in off-diagonal and diagonal models
have markedly different spectra, i.e., the former has a
Dysonian singularity \cite{ziman7,eggarter} at the band center while the
latter has a power-law singularity at the band edge. Hence the short-range
correlations are expected to influence off-diagonal spectra more than in the
diagonal case.
\setcounter{equation}{0}
\section{Results}
The distribution and localization of the exciton modes are characterized by
the Lyapunov exponent. In the case of one-dimensional arrays with
nearest-neighbor interactions, the Lyapunov exponent can be determined by
making use of mode-counting techniques \cite{dean,hub73}. Recently we
studied \cite{avh05} similar excitons with orientationally disordered
couplings in conjugated polymers and calculated the full spectrum and
optical line shapes. The goal here is to look at the effects of
exponentially correlated disorder on the behavior of the $IDOS$ and $ILL$
near the center of the band ($E=0$). We have produced three sets of data for
each case varying the correlation length i.e., c) and the
strength of the disorder, $A,$ with $A=0.25,0.50,0.75 $.

\begin{figure}
\unitlength=1cm
\begin{picture}(15,5)
\unitlength=1mm
\centerline{
\epsfysize=7.5cm
\epsfbox{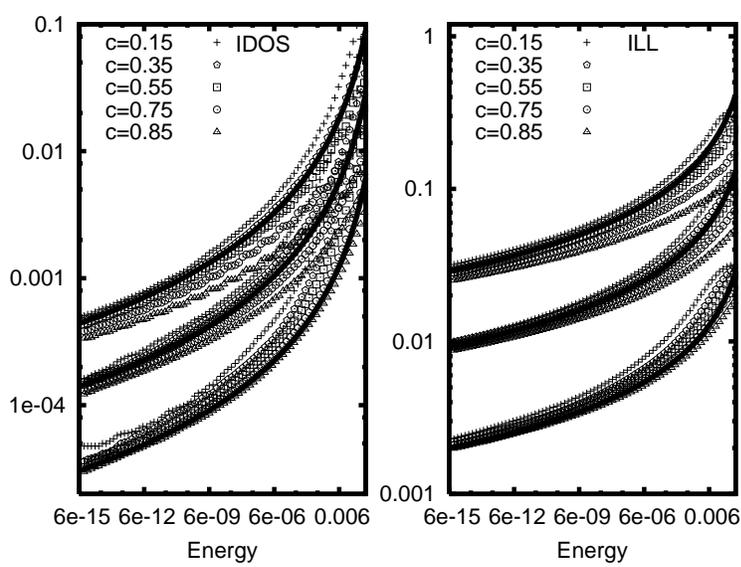}
}
\end{picture}
\caption{The integrated density of states ($IDOS$, left panel) and the
inverse localization length ($ILL$, right panel) plots for the correlated
distributions of the dipolar type couplings with $A=0.25,0.50, 0.75$ bottom
to top.  Symbols are data for a chain of $10^{8}$ sites for respective
concentrations divided by $f(c)=\frac{c}{1-c}$.
The lines, shown as guides to the eye, are plots of the amplitude function
$f(0.5)(\sqrt 2 \tanh^{-1}A)^2$ divided by $\ln(E^2)$ for the $ILL$
or its square for the $IDOS$.
The parameter $c$ is related to the correlation
length $l(c)$ by the equation $l(c) = - 1/\ln|1-2c|$.}
\label{fig:ids0}
\end{figure}

Figure~\ref{fig:ids0} displays the results of the $ILL$ and $IDOS$
for exponentially correlated distributions of the couplings.
For fixed $A$, we divided the data produced for a particular
concentration $c$ by a scaling factor $f(c) = c/(1-c)$ 
(see next section and the appendix).
The data for selected concentrations are shown in Fig.~\ref{fig:ids0}
for the indicated values of $A$. 
We expect first that for a range of concentrations
the data should collapse to a Dysonian singularity for fixed $A$.
However, Fig.~\ref{fig:ids0} reveals that as $c$ increases,
the $ILL$ and the $IDOS$ take values that differ increasingly from
predictions based on Ziman's analysis \cite{ziman7} and a slow approach to
the band center is observed. In this limit, the data clearly show
qualitative scaling behavior but deviations with varying magnitude exist as $%
c\rightarrow 1$ and $c\rightarrow 0$ where the correlation length is
infinite. Notice also that the deviations are stronger for the $IDOS$ than
for the $ILL$.

\begin{figure}
\label{fig:idsf}
\unitlength=1cm
\begin{picture}(15,5)
\unitlength=1mm
\centerline{
\epsfysize=7.5cm
\epsfbox{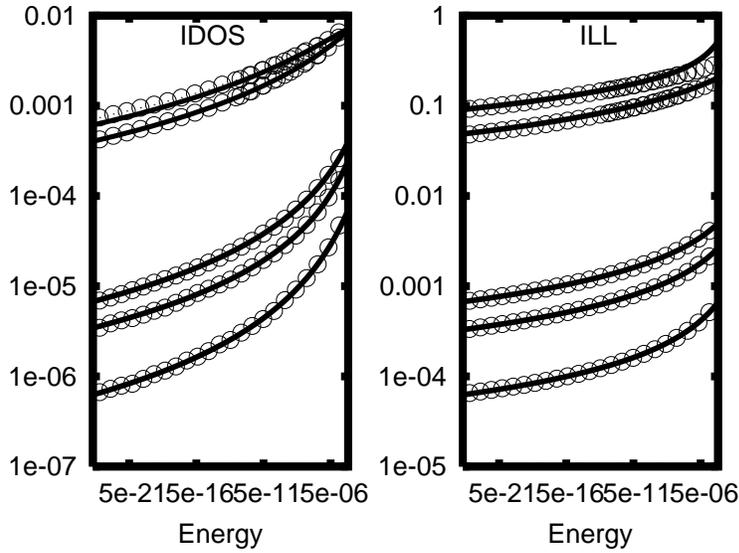}
}
\end{picture}
\caption{The integrated density of states ($IDOS$, left panel) and the
inverse localization length ($ILL$, right panel) plots for the correlated
distributions of the dipolar type couplings with $A=0.50$. Symbols are data
for a chain of $10^{8}$ sites for particular concentrations (where
significant deviations occur) $c=0.01,0.05,0.1,0.9, 0.95$ from bottom the
top and the lines are the scaling function $f(c)(\sqrt 2
\tanh^{-1}0.5)^2$ divided by $\ln(E^2)$ for the $ILL$ or
its square for the $IDOS$ plus the higher terms given
in Eqs.\ (\ref{idosc})--(
\ref{illc}) where $f(c)=\frac{c}{1-c}$.
The parameter $c$ is related to the correlation length $l(c)$
by the equation $l(c) = - 1/\ln|1-2c|$.}
\end{figure}

\noindent Clearly as energy decreases, the scaling gets better; however, the
deviations become worse as one gets closer to $c\rightarrow 0,1$. To
see whether longer chains and/or lower energies improve the situation, we
checked further and found that the longer chain did not improve the
agreement appreciably, but as the energy is lowered, the results follow the
scaling behavior more closely up to a certain limiting $c$. Similar
disagreement between the data and the Dysonian (with a single term)
singularity were encountered in the SUSY \cite{jp1} off-diagonal tight
binding case for continuous exponential correlations. The analytical
investigations \cite{jp2} revealed that the spectra for the exponential
correlation can be expanded for the $IDOS$ as

\begin{equation}  \label{idosc}
IDOS-{\frac{1 }{2}}=a_1/|\ln(E^2)|^2 + a_2/|ln(E^2)|^3 + a_3/|ln(E^2)|^4
+\dots,
\end{equation}

\noindent and for the inverse localization length ($ILL$)

\begin{equation}  \label{illc}
ILL=b_1/|ln(E^2)|+b_2/|ln(E^2)|^2+b_3/|ln(E^2)|^3 \dots ,
\end{equation}

\noindent where $a_{1},a_{2},a_{3},b_{1},b_{2}$ are given as of functions of
the correlation length multiplied by the strength \cite{jp1,jp2} of the
exponential correlations. The equations\ (\ref{idos})--(\ref{ill}) are only
the first order term in the expansion and $a_{1}$ and $b_{1}$ should
correspond to correlated amplitude factor $V^C(c)$. Figure~\ref{fig:idsf}
clearly shows the improved results obtained using the
Eqs.\ (\ref{idosc})--(\ref{illc})
for the selected concentrations where the
worst deviations followed from the first order result: $%
c=0.01,0.05,0.1,0.9,0.95$. Unfortunately, for $c<0.01$ and $c>0.95$, this
type of expansion appears to fail; however,
other forms of fitting with fractional
powers can give better results such as
tried in the SUSY case \cite{jp1,jp2}. We will not here
discus significance of this fractional fitting since
there doesn't seem to be any analytic theory that predicts it.

\setcounter{equation}{0}
\section{Discussion}
To shed some light on the scaling behavior for the correlated case, we first
compute the binary uncorrelated case. A simple calculation of the
logarithmic variance, the amplitude of the spectrum, for the uncorrelated
case given in Eqs.\ (\ref{idos})--(\ref{ill}) yields the result

\begin{equation}
V^U(c)=4c(1-c){\left( \sqrt 2 \tanh ^{-1}A\right) }^{2}.  \label{uvar}
\end{equation}

\begin{figure}
\label{fig:scl}
\unitlength=1cm
\begin{picture}(15,5)
\unitlength=1mm
\centerline{
\epsfysize=7.5cm
\epsfbox{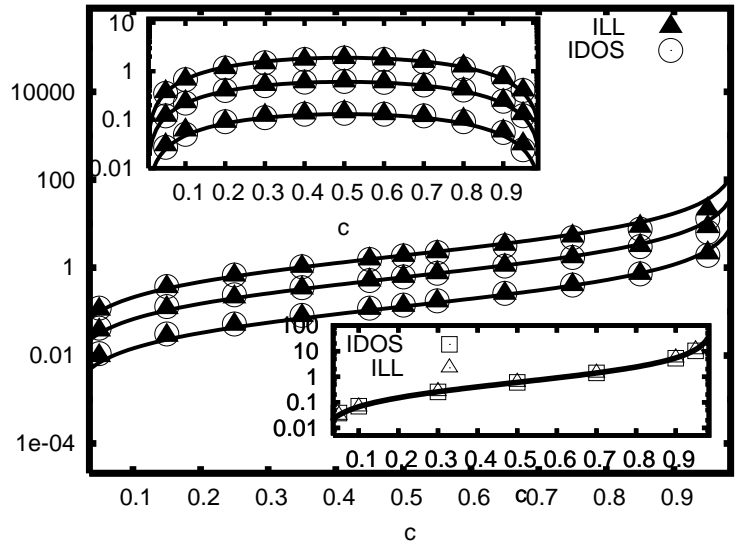}
}
\end{picture}
\caption{The asymptotic amplitude factors for
correlated disorder vs concentration $c$
with $A=0.25,0.50, 0.75$ bottom to top.
In the main plot, the lines show the correlated amplitude
$V^C(c)=f(c){\left (\sqrt 2 \tanh^{-1} A \right )}^2$
where $f(c)=\frac{c}{1-c}$.
The top inset is the same plot for
uncorrelated case for the same $A$ values.
The lines are a plots of the uncorrelated amplitude $V^U(c)=g(c){\left
(\sqrt 2 \tanh^{-1} A \right )}^2$ where $g(c)=4c(1-c)$.
The bottom inset shows
$f(c){ \left ( \sqrt 2 \tanh^{-1} A \right )}^2$,
the expansion coefficient of the $IDOS$ and the
$ILL$ $a_1$ and $b_1$ , respectively vs $c$ for $A=0.5$.  The parameter
$c$ is related to the correlation length $l(c)$
by the equation $l(c) = - 1/\ln|1-2c|$.}
\end{figure}

\noindent The amplitude factor for the uncorrelated disorder clearly
displays a scaling behavior in concentration $c$ as well as in $A$. We can
identify the scaling functions for the concentration $g(c)=4c(1-c)$ and the
strength of the deviation $h(A)=(\sqrt{2}\tanh ^{-1}A)^{2}$. The top inset
in Fig.~\ref{fig:scl} displays the uncorrelated data vs concentrations. The
three lines are the calculated amplitudes $V^{U}(c)=h(A)g(c)$ presented in
Eq.\ (\ref{uvar}) for $A=0.25,0.50,0.75$. As expected, the theory for the
uncorrelated case works well

The amplitude factors in Fig.~\ref{fig:scl} (large panel) are found from the
best fit to the Dysonian singularity. They are obtained matching the
smallest value of the energy rather than the value obtained by fitting over
a range of energies. In this way, we will be more likely to get a value
characterizing the asymptotic region. The calculated amplitude scaling
function is also shown in Fig.~\ref{fig:scl} for the correlated case $%
V^C(c)=c(1-c)^{-1}{\left( \sqrt{2}\tanh ^{-1}A\right) }^{2}$ for the three
specified $A$ values above. In the appendix we will argue why this formula
holds. We notice immediately that the concentration dependence has a
different form $f(c)=\frac{c}{1-c}$ as compare to $g(c)=4c(1-c)$ \ of the
uncorrelated case while the scaling on $A$ is the same for both cases. $h(A)=%
{\left( \sqrt{2}\tanh ^{-1}A\right) }^{2}$. This asymptotic fit can further
be checked by plotting the $a_{1}$ and $b_{1}$ in the expansion above since
they correspond the coefficient of the Dysonian singularity. The bottom
inset displays the $a_{1}$ and $b_{1}$ obtained by the best fit as compare
to the theoretical line $V^C(c)$ (for $A=0.5$). Both plots clearly support
the asymptotic amplitude scaling.

Although the results obtained by asymptotic fit are clearly reproduced by
the theoretical amplitude factor, there are striking differences between
correlated and uncorrelated cases. What is particularly interesting and not
expected is that the results for correlated disorder is not symmetric about
$c=0.5$ like those of the uncorrelated case. Also, from our scaling analysis
the $ILL$ coefficient becomes large as $c\rightarrow 1$, but vanishes as $%
c\rightarrow 0$. This is surprising since the modes for $c=1$ are extended
just like the modes for $c=0$. The divergence of the amplitude factor as $%
c\rightarrow 1,$ is not inconsistent with the results for $c=1$ since the
energy interval over which scaling holds shrinks to zero in the same limit.
This can be seen from Eqs.\ (\ref{idos})--(\ref{ill}) by setting either the $%
IDOS$ or the $ILL$ to a constant value and solving for the cut-off energy as
as a function of $c$.
As shown in the appendix, the amplitude factors perform a 
correlated random walk in parameter space. For our case, the square of 
the length of the random walk is proportional to $cN/(1-c)$ not $N$ 
alone as in the uncorrelated case. That is the source of the marked 
difference between the two cases. As the concentration approaches $1$, 
divergent behavior occurs. However, we note that we cannot take c 
arbitrarily close to $1$; beyond a certain point, the random walk sequence 
can not be constructed \cite{eggarter}.

\setcounter{equation}{0}
\section{Summary}

We have investigated the dynamics of 1D Frenkel exciton systems with
correlated off-diagonal disorder. We have used a negative eigenvalue
counting technique \cite{dean} which provides a simple and physically
transparent analysis of the $IDOS$ and the $ILL$. We investigated the
question of scaling when there was correlated off-diagonal disorder such as
might occur, for example, when the off-diagonal interaction depended on the
distance between two ions and thus would be affected by small displacements
of the ions from their equilibrium positions \cite{dip}. Our numerical data
indicate that in the correlated case the scaling behavior found by Ziman
\cite{ziman7} is followed only for a limited range of $c$ and very small $E$
. We compared our results with the scaling predictions of Ziman \cite{ziman7}
and with the similar tight binding correlated electronic case \cite{jp1,jp2}%
. We computed the asymptotic amplitude factors
as a function of concentration and found
that they are equal to the uncorrelated variance at $c=0.5$ multiplied by
the factor $\frac{c}{1-c}$ (see appendix). This factor \cite{avgin93,av05}
has played a very interesting scaling role in number of unrelated problems.
We found that with the new variance for correlated distributions rather
unexpected behavior was obtained as compare to that of the uncorrelated
case. In particular, the asymptotic amplitude
factor is not symmetric about $c=0.5$
while the $ILL$ diverges for concentrations approaching $1$ in contrast to
what is observed in the uncorrelated case. This amplitude factor can explain
only the first order expansion in a Dyson-type singularity as supported by
the SUSY model \cite{jp1,jp2}. As shown in the appendix, if the random walk
observation of Eggarter \emph{et}$\,$\emph{al. is} implemented, with
correlation, the obtained amplitude factor can reproduce the data within a
shrinking energy range as $E\rightarrow 0$ as $c$ is increased. Finally, we
should emphasize that our work is a numerical study supplemented by the
approximate theoretical analysis outlined in the Appendix. As mentioned
previously, the surprising result is the discovery that near the center of
the exciton band, a model with correlated disorder showed asymptotic
behavior that is
similar to the behavior of systems without correlations, the only
difference being in the amplitude factor appearing in the limiting
expressions for the IDOS and ILL. Like those of essentially all numerical
studies, our results are approximate. We hope that the availability of the
numerical findings together with our analytical results will stimulate
rigorous analyses of the model that may shed light on the origin of the
similarity.

\setcounter{equation}{0}
\section{Acknowledgement}
This work is partially sponsored by the Scientific and Technical Research
Council of Turkey (TUBITAK) and Ege university research Grant (BAP).

\setcounter{equation}{0}  
\section{Appendix}
In this section we present the amplitude scaling related to the first order
expansion coefficients $a_{1}$ and $b_{1}$ in Eqs.\ (\ref{idosc})--(\ref
{illc}). Similar scaling but in different cases has appeared before \cite
{rus,stinch,westerberg,avgin93,av05}. However the off-diagonal problem here is
more involved than those cases. First the Lyapunov exponent at the band
center $E=0$ is calculated using Eq.\ (\ref{tb}) that can be rearranged \cite
{irn,ziman7,eggarter}

\begin{equation}
\gamma(0)=\frac {1}{N}\ln(U_N/U_1)=\frac {1}{N} \sum_{n=1}%
\ln(U_{2n+1}/U_{2n})= \frac{1}{N}\sum_{n=1}\ln(J_{2n+1}/J_{2n})+\frac {i \pi
}{2}.  \label{l1}
\end{equation}

\noindent The bipartite nature (the chiral symmetry) of the Eq.\ (\ref{tb})
is responsible for the logarithmic variance as rigorously shown in \cite
{inui} and eventually the Dyson- type singularity \cite{eggarter} since $
\gamma (0)$ executes a random walk with a step $\Delta _{n}=\ln
(J_{2n+1})-\ln (J_{2n})$. For the uncorrelated case the averages yield $
<\Delta _{n}>=0$ and $<{\Delta _{n}}^{2}>=\frac{1}{2}[{<(\ln (J^{2}))^{2}>-{
<\ln (J^{2})>}^{2}}]$. The Lyapunov exponent \cite{eggarter,inui} takes this
form $\gamma (0)=\frac{1}{N}\sqrt{<(\sum_{n=1}\Delta _{n})^{2}>}+\frac{i\pi
}{2}$. The imaginary part here $IDOS=\frac{\Im {\gamma (0)}}{\pi }=0.5$ and
the real part is related to the $ILL$. To make our point, we can attack the
problem with a different angle. The average of the summation $
<\sum_{n=1}\Delta _{n}>=0$ for both correlated and uncorrelated cases;
whereas, the average of the squared summation $\sum_{n=1}<{\Delta _{n}}
^{2}>+2\sum_{m\not=n}<\Delta _{n}\Delta _{m}>$. It further reduces to $N<{
\Delta }^{2}>$ since the second sum is zero for the independent random
numbers but non zero for the correlated random numbers. For the correlated
disorder the summation is not easy. The complication arises when correlation
present since $<\ln (J_{n})>$ is not equal to its ensemble average rather it
dependents on $n$. But this complication can be avoided in a following way.
Let us calculate\ the Taylor expansion of $\ln (J_{n})=\ln (1+A\xi _{n})$
which reads
\begin{equation}
\ln(1+A\xi_n)=A\xi_n-(A\xi_n)^2/2+(A\xi_n)^3/3- (A\xi_n)^4/4+(A\xi_n)^5/5+
\dots.  \label{a1}
\end{equation}

\noindent As ${\xi _{n}}^{2k}=1$ but
for odd powers ${\xi _{n}}^{2k+1}=\xi _{n}$. Arranging even and odd powers
we get

\begin{equation}
\ln(1+A\xi_n)=(A+A^3/3+A^5/5+A^7/7+ \dots) \xi_n-(A^2/2+A^4/4+A^6/6+ \dots) .
\label{a2}
\end{equation}

\noindent The first sum (with odd powers of A) is just the expansion of $
tanh^{-1}A$ and let the second sum (with even powers) be $D$ which will be
eliminated when it is inserted in $\Delta _{n}=tanh^{-1}A(\xi _{2n+1}-\xi
_{2n})$. The same expansion can be used to calculate the uncorrelated case
as well; the step takes the form $\Delta _{n}=tanh^{-1}A(x_{2n+1}-x_{2n})$
(recall that number of steps $N/2$). Using Eq.\ (\ref{dist}), the variance
is given by $V^{U}(c)=<\frac{2}{N}\sum_{n}{\Delta _{n}}^{2}>=2<{\Delta }
^{2}>=4c(1-c)(\sqrt{2}tanh^{-1}A)^{2}$ so that the $ILL$ becomes $\Re {
\gamma (0)}=\sqrt{V^{U}(c)/N}=\sqrt{2}tanh^{-1}A\sqrt{4c(1-c)/N}$. For the
correlated case, we have $V^{C}(c)=\frac{2}{N}<(\sum_{n}\Delta
_{n})^{2}>=(tanh^{-1}A)^{2}<(\sum_{n}(\xi _{2n+1}-\xi _{2n}))^{2}>$. Taking
the square, we get three terms $<(\sum \xi _{odd/even})^{2}>$ and twice $
<\sum \xi _{odd}\sum \xi _{even}>$, hence the summation results in $N\frac{c
}{1-c}$ where inverse of this factor is encountered in previous works \cite
{av05,avgin93,rus,westerberg}. The correlated random walk results in $V^{C}(c)=%
\frac{c}{1-c}(\sqrt{2}tanh^{-1}A)^{2}$ and the $ILL=\Re \gamma (0)=\sqrt{
V^{C}(c)/N}=\sqrt{2}tanh^{-1}A\sqrt{\frac{c}{(1-c)N}}$. In contrast, the $
IDOS$ value at the band center is not affected by the correlations since the
distribution in eigenvalues is symmetric about the mid-point. \ The
calculation above holds at the band center; however Eggarter \emph{et}$\,$
\emph{al.}, \cite{eggarter} have shown that for a certain range of $E$ close
to the band center, the random walk behavior still held and the $ILL$ and $
IDOS$ can be represented as in Eqs.\ (\ref{idosc})--(\ref{illc}) where the
amplitude factor is the same as calculated at the band center.



\begin{thebibliography}{99}
\bibitem{dip} 
According to the Frenkel exciton model, 
the transfer integral usual dipole-dipole 
interaction has the inverse cube of the distance between the dipoles. 
Offdiagonal transfer integral disorder is due to the distribution 
of bond length. Assuming the transition dipoles, the intersite transfer
dipole-dipole interaction is given as $J_{n,m}={\frac{1
} {{|(a_0n+\delta_n)-(a_0m+\delta_m)|}^{3}}}$ where $\delta_n$ is 
the positional deviation from their equilibrium 
position $a_0n$ of transition dipoles. The nearest distance between the atoms
is $a_0$ and $n$ is the site index.  Since the 
interaction decreases rapidly over the $n$ to $m$ distance, 
regarding nearest-neighbor couplings is sufficient as a first approximation.  
Assuming unit neighbor distance the tranfer integral reduces to 
$J_{n,n+1}={\frac{1} {{|1-B\xi _{n}|}^{3}}}$ 
where $B$ is the strength and $\xi_n=(\delta_{n+1}-\delta_n)/B$ is the
random sign of the deviation from the regular lattice point. 
Notice that $J_{n,n+1}$ is used in the text as $J_{n}$.
The deviations from the equilibrium positions introduces a
correlation.  If $B$ is small, one can expand the coupling 
as $J_{n} \simeq 1+3B\xi _{n} $. Since B
is a free parameter, one can replace 3B with A without loss of generality.
Note that using the approximate form in place of the cubic form does not
make an essential change in the spectra apart from a numerical factor since
the spectra are related to averages of the logarithm, $<\ln({J_{n}}^{2})>$
and our numerical study confirmed this.
\end{thebibliography}
\end{document}